\documentclass[twocolumn,secnumarabic,amssymb,aps, prb,superscriptaddress]{revtex4-2}

\usepackage{hyperref}

\newcommand{\revise}[1]{\textcolor{black}{#1}}

\usepackage{amsmath,graphicx,subfigure}
\usepackage{newtxtext,newtxmath}
\begin{document}

\title{Integral Variable Range Hopping for Modeling Electrical Transport in Disordered Systems}%

\author{Chenxin Qin}
\affiliation{School of Physics, Peking University, Beijing 100871, People's Republic of China}

\author{Chenyan Wang}
\affiliation{School of Physics, Peking University, Beijing 100871, People's Republic of China}

\author{Mouyang Cheng}
\email{vipandyc@alumni.pku.edu.cn}
\affiliation{School of Physics, Peking University, Beijing 100871, People's Republic of China}

\author{Ji Chen}
\email{ji.chen@pku.edu.cn}
\affiliation{School of Physics, Peking University, Beijing 100871, People's Republic of China}
\affiliation{Interdisciplinary Institute of Light-Element Quantum Materials and Research Center for Light-Element Advanced Materials, Peking University, Beijing 100871, People's Republic of China}
\affiliation{State Key Laboratory of Artificial Microstructure and Mesoscopic Physics and Frontiers Science Center for Nano-Optoelectronics, Peking University, Beijing 100871, People's Republic of China}

\begin{abstract}
    The variable range hopping (VRH) model has been widely applied to describe electrical transport in disordered systems, providing theoretical formulas to fit temperature-dependent electric conductivity. 
    These models rely on oversimplified assumptions that restrict their applicability and result in problematic fitting behaviors, yet their overusing situation is becoming increasingly serious.
    In this work we formulate an integral variable range hopping (IVRH) model, which replaces the empirical temperature power-law dependence in standard VRH theories with a physics-inspired integral formulation. 
    The model builds upon the standard hopping probability $\omega(R)$ w.r.t. hopping distance $R$ and incorporates the density of accessible electronic states through an effective volume function $V(R)$, which reflects the influence of system geometry. The IVRH formulation inherently reproduces both the Mott behavior at low temperatures and the Arrhenius behavior at high temperatures, respectively, and enables a smooth transition between the two regimes. 
    We apply the IVRH model to two-dimensional, three-dimensional, and multi-layered systems. Monte Carlo simulations validate the model’s predictions and yield consistent values for the fitting parameters, with substantially reduced variances compared to fitting using the standard VRH model. 
    Furthermore, the improved robustness of IVRH also extends to the transport measurements in monolayer MoS$_2$ system and monolayer WS$_2$ system, enabling more physically meaningful interpretation.
    IVRH model offers a more stable and physically sound framework for interpreting hopping transport in low-dimensional amorphous materials, providing deeper insights into the universal geometric scaling factors that govern charge transport in disordered systems.
\end{abstract}

\maketitle

\section{Introduction}
Disordered materials and disorders in materials provide versatile platforms to extend their applications in electronic devices \cite{huang2013imaging,toh2020synthesis,bai2024nitrogen,liu2025amorphous,yao2025transferrable,li2025mechanical,tian2023disorder,cheng2025predicting}.
The disordered electronic structures and disorder-induced transport phenomena have been extensively investigated across amorphous semiconductors, oxides, and low-dimensional carbon systems, revealing rich metal-insulator transitions and localization physics \cite{deringer2021origins,prasai2012properties,van2012insulating,kapko2010electronic,thapa2022ab,antidormi2022emerging,imada1998metal}.
To understand the transport physics of disordered systems, notable theoretical contributions were made by Anderson and Mott \cite{anderson1958absence,mott1967electrons,mott1968conduction,mott2012electronic}, who shared the 1977 Nobel Prize in physics with Van Vleck. 
Despite the simplicity, these models have captured the main features of charge transport in disordered systems \cite{tian2023disorder,cheng2024regulating,bhattacharjee2025anderson,bx6m-635y,ortuno2022numerical,jiang2025variable,tsigankov2002variable,kinkhabwala2006numerical}.
In particular, with experimentally measured temperature-dependent conductivity, Mott's variable range hopping (VRH) model remains the most widely used theory to judge the role of disorder in experiment.

However, despite the broad applicability of the VRH model, challenges and controversy persist regarding its assumptions and reliability in complex disordered systems.
In the original VRH model, conduction occurs via thermally activated tunneling between localized states, predicting $\log \sigma \propto T^{-1/(d+1)}$ in $d$-dimensional systems \cite{mott1968conduction}.
Later, Efros and Shklovskii extended the model to account for Coulomb interactions and the new model would give a universal $\log \sigma \propto T^{-1/2}$ behavior \cite{ALEfros_1975}.
Both of these two laws are observed in experiment \cite{PhysRevB.48.2312,PhysRevB.90.054204,doi:10.1021/nn101376u}; however, due to the limited range of experimental data and the close resemblance between these functional forms, distinguishing between these two models remains challenging \cite{PhysRevB.86.235423}.
Moreover, a range of experimental systems has observed that VRH only occurs at lower temperatures, while the temperature dependence of conductivity at high temperatures remains thermal-activated \cite{PhysRevLett.92.216802,PhysRevLett.104.056801,qiu2013hopping}, and the transition from thermal-activated regime to VRH regime remains unclear. 
Although such a transition has been empirically proposed \cite{OKUTAN2005176}, and recently reproduced by a random resistor network research \cite{bx6m-635y}, a unified and quantitatively grounded theoretical description is still lacking. 
Besides, Mott's VRH model is not suitable for multi-layer systems, where the interlayer hopping is not negligible and largely limited by the finite height, thus the vertical boundary is no longer periodic.

In order to provide the VRH with stable and physically grounded description, we propose an integral variable range hopping (IVRH) model, which replaces the empirical temperature exponent of VRH with an integral formulation based on slightly different physical assumptions. The key idea is to express the hopping probability ($\omega$) as a function of both hopping distance $R$ and energy cost $\Delta E$, and to introduce an effective volume function ($V(R)$) that captures the number of accessible states. 
Then, we eliminate the $R$ dependence and obtain the conductivity as a function of temperature by integrating over $R$. This approach naturally recovers both Arrhenius and Mott-like behavior, and also extends straightforwardly to multi-layer systems, where the hopping volume $V(R)$ is modified by out-of-plane confinement and includes the dependence of the number of layers. 

\begin{figure*}[htpb]
    \centering
    \includegraphics[width=0.7\linewidth]{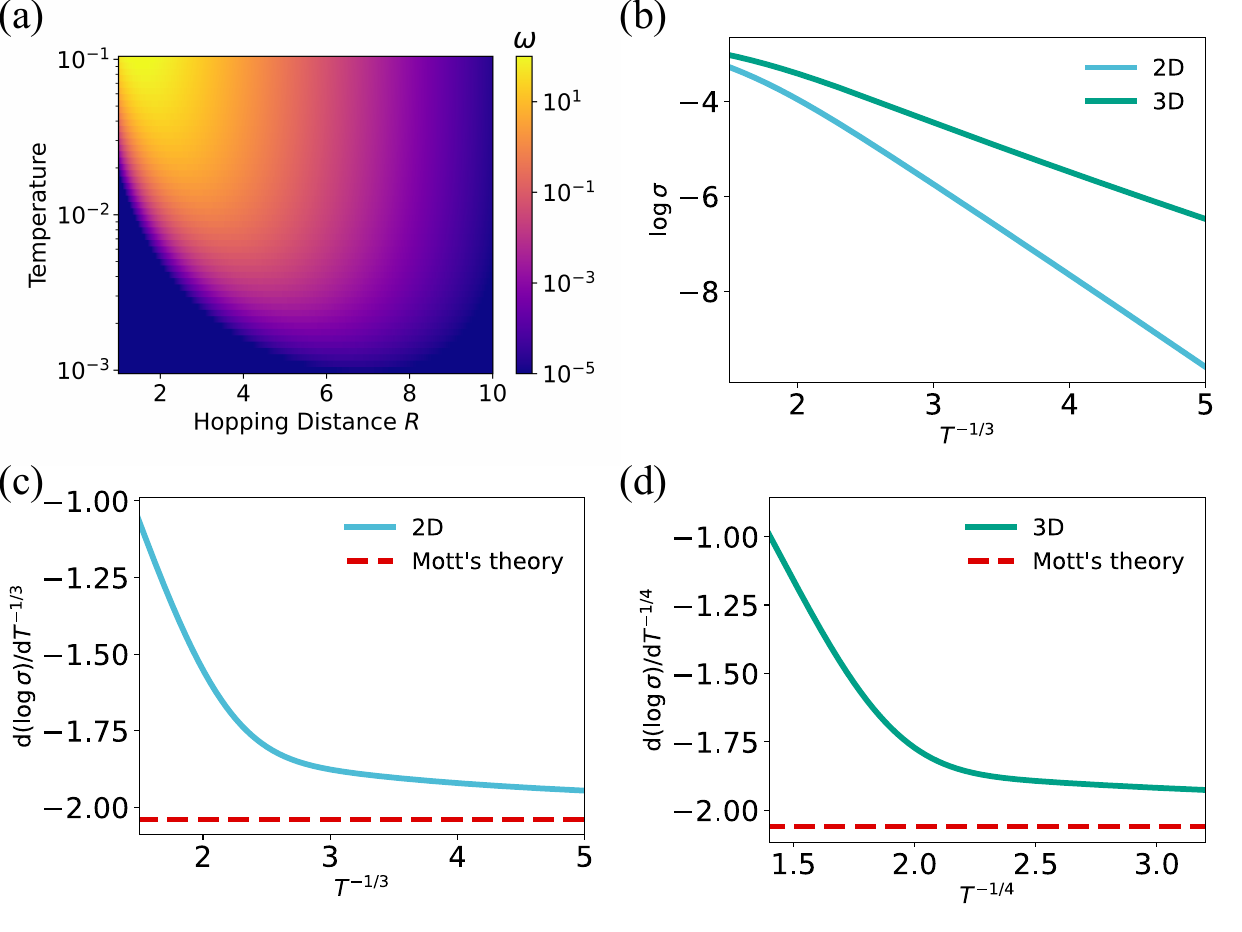}
   
    \caption{IVRH in 2D and 3D systems and comparison with Mott's VRH model. \textbf{(a)} Dependence of the hopping probability $\omega$ on the hopping radius $R$ at different temperatures in 2D systems. From top to bottom, the temperatures of different curves decrease. \textbf{(b)} The temperature dependence of the IVRH conductivity in 2D and 3D systems. \textbf{(c,d)} First derivatives of the $\sigma(T)$ curves in 2D systems \textbf{(c)} and 3D systems \textbf{(d)} w.r.t. temperature $T$, respectively. The solid line corresponds to the IVRH model, and the red dashed line corresponds to Mott’s law, which is independent of $T$. Parameters used: $\alpha=\beta=D_0=k_B=1$ as defined in Eq.\,\ref{e2}.}
    \label{fig1}
\end{figure*}

To justify our formulation of IVRH, we validate our theoretical model using Monte Carlo simulations. The extracted IVRH parameter varies systematically with dimensionality and shows significantly lower fitting uncertainty compared to conventional Mott exponents. We also test the IVRH model with electrical transport experimental results in monolayer transition metal dichalcogenide systems, and our model successfully covers the observations and digs out more information, such as the transition from thermal-activated behavior to VRH behavior and the gate-tuned delocalization. These results establish IVRH as a stable and versatile framework for analyzing hopping conduction in amorphous and layered materials. Given its generality and consistency, our IVRH model offers a unified viewpoint for studying electrical hopping transport across a wide range of low-dimensional disordered systems.

\section{Theory}
An essential point in the VRH model is the relation of the electron hopping probability $\omega$ w.r.t. hopping distance $R$:
\begin{equation}
\omega \propto \exp\left(-2\alpha R - \frac{\Delta E}{k_BT}\right), \label{e1}
\end{equation}
where $\alpha$ is the inverse localization length. When the hopping probability is independent of hopping radius, it recovers the purely thermal-activated Arrhenius law, where the electric conductivity scales as $\sigma\propto \exp(-\Delta E/k_BT)$. In the general VRH regime, the energy cost $\Delta E$ is related to the requirement that an accessible state exists within the hopping volume, defined as $V(R)$. For example, in three dimensions, this condition gives $\frac{4}{3}\pi R^3 D_0 \Delta E \sim 1$, where $D_0$ is a constant that approximates the density of states near the Fermi level. This leads to a modified hopping probability:
\begin{equation}
\omega \propto \exp\left(-2\alpha R - \frac{\beta}{V(R) D_0 k_BT}\right), \label{e2}
\end{equation}
where $\beta$ is an undetermined constant. This expression encapsulates the fundamental trade-off in VRH: longer hops reduce the energy penalty but are suppressed by the tunneling factor, while shorter hops are more probable but access fewer available states. Fig.\ref{fig1}(a) shows the $\omega(R)$ at different temperatures. With the temperature decreasing (from top to bottom), the thermally assisted hopping becomes increasingly suppressed. 
Charge transport is then dominated by hopping over longer distances to energetically closer localized states, reducing the activation energy at the cost of increased spatial separation. This tradeoff between spatial and energetic penalties underlies the mechanism of VRH.

Notably, in the original VRH model, Mott further assumes that the conductivity is proportional to the maximum of the function $\omega(R)$.
Via the extremum condition, Mott analytically solves the most probable hopping distance (with $\beta=1$):
\begin{equation}
    R^* = \arg\max_{R}\,\omega(R)=\left\{
    \begin{aligned}
        \left(\frac{1}{\pi D_0\alpha k_BT}\right)^{\frac{1}{3}},\text{2D system}\\
        \left(\frac{8}{9\pi D_0\alpha k_BT}\right)^{\frac{1}{4}},\text{3D system}
    \end{aligned}
    \right.
\end{equation}
 and then concludes that the maximum of the function $\omega(R^*)$, hence the conductivity, is proportional to $T^{-\frac{1}{d+1}}$.
\begin{figure*}[t]
    \centering
    \includegraphics[width=0.85\linewidth]{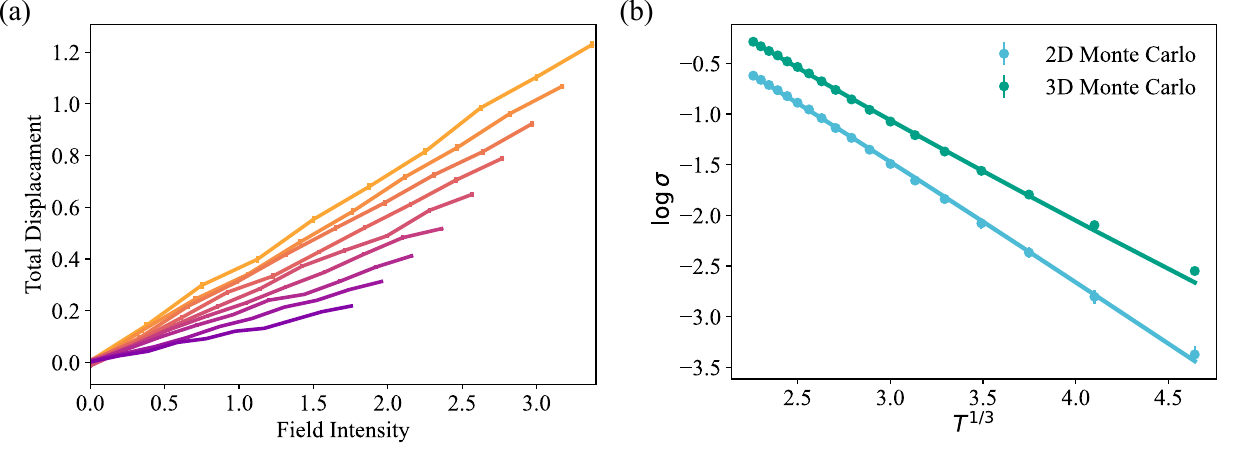}
    \caption{Monte Carlo simulation and conductivity extraction on 2D and 3D systems. \textbf{(a)} The relationship between total displacement and field intensity. The slope is proportional to conductivity. From bottom to top, the temperature $T$= 0.01, 0.0145, 0.019,  0.0235, 0.028,  0.0325, 0.037,  0.0415, 0.046,  0.0505, respectively. \textbf{(b)} Monte Carlo results (dots) and corresponding IVRH model fits (curves) for 2D (blue) and 3D (green) systems. Parameters used in the simulation: $\alpha = 0.5$, $D_0 = 1$ as defined in Eq.\,\ref{e2}. }
    \label{fig2}
\end{figure*}

Here, instead of solving for a single optimal hopping distance, we numerically integrate Eq.\eqref{e2} over all possible hopping distances $R$ as follows:
\begin{equation}
    \sigma=A\int_{R_0}^\infty\exp\left(-2\alpha R - \frac{\beta}{V(R) D_0 T}\right)\mathrm{d}R\label{eq4}
\end{equation} 
A lower bound $R_0$ is imposed to reflect the minimum separation between localized sites. This integration is based on the physical intuition that various hopping distances contribute to the conductivity, instead of only the most probable distance. Fig.\ref{fig1}(b) shows the theoretical results for 2D and 3D systems. At high temperature, due to the existence of a minimum hopping distance $R_0$, IVRH approximates to Arrhenius' law.  
At low temperature, i.e. larger $T^{-1/3}$, the $\log\sigma-T^{-1/3}$ curve exhibits linear behavior in $\log\sigma-T^{-1/3}$ scale, which is similar to Mott's law. However, as shown in Figs.\ref{fig1}(c)(d), the derivative of the IVRH curves asymptotically approaches the constant results of Mott's theory at low temperatures but retains curvature\cite{PhysRevB.4.2612}.

Furthermore, when using the IVRH to fit complex experimental data, additional numerical techniques can be introduced to stabilize the fitting.
First, a variation transformation with $u=1/R$ to convert the infinite integral to a finite one:
\begin{equation}
    \log\sigma=A+B\log\left[\int^{1/R_0}_0\frac{\exp\left(-2\alpha'/u  - \frac{\beta}{V(1/u) T}\right)}{u^2}\mathrm{d}u\right]
\end{equation}
where $B=1/D_0,$ $\alpha'=\alpha D_0$. 
Second, the numerical integration error can be systematically reduced using Simpson’s rule by decreasing the spacing between integration points 
In this study, we adopt the Fortran library QUADPACK\cite{piessens1983quadpack} to realize the numerical integral above.

\section{Results}

\subsection{Model validation with Monte Carlo simulation}
To connect the IVRH model with experimental observations, it is essential to determine the parameter $\beta$ in Eq.\eqref{e2}. Following the methodology of previous studies on hopping transport in disordered systems \cite{tsigankov2002variable,kinkhabwala2006numerical}, we perform Monte Carlo simulations based on a lattice model governed by the Hamiltonian:
\begin{equation}
    H=\sum_i E_in_i
\end{equation}
where $E_i$ are random site energies uniformly distributed within [-$W/2,W/2$], and $n_i$ is the occupation number. The simulations are carried out in a three-dimensional lattice of size $L \times L \times H$ with periodic boundary conditions depending on specific systems. 

To simulate the hopping process, we randomly select pairs of lattice sites. Each pair is added to the candidate set with a probability proportional to $\exp(-2\alpha R)$. In order to calculate the conductivity, we add a small electric field $F$ along the $x$ axis as a perturbation. Choosing a pair from the candidate set, the energy difference associated with a hop from site $i$ to site $j$ is given by $\Delta E = E_j - E_i - F\Delta x$, where $\Delta x$ is the displacement in the field direction. The hopping event is accepted with probability $1/[\exp(\Delta E/T)+1]$. Each accepted trial move contributes to the net displacement along the $x$-axis. After a number of simulation steps $N$, the total displacement $X_{\text{total}}$ is recorded. When $F/\alpha$ is much smaller compared with $k_BT$ and $W$, $X_{\text{total}}$ grows linearly as $F$ increases, thus we can extract conductivity from
\begin{equation}
    \sigma F\sim\frac{X_{\text{total}}}{N}
\end{equation}

\begin{figure*}[]
    \centering
    \includegraphics[width=0.75\linewidth]{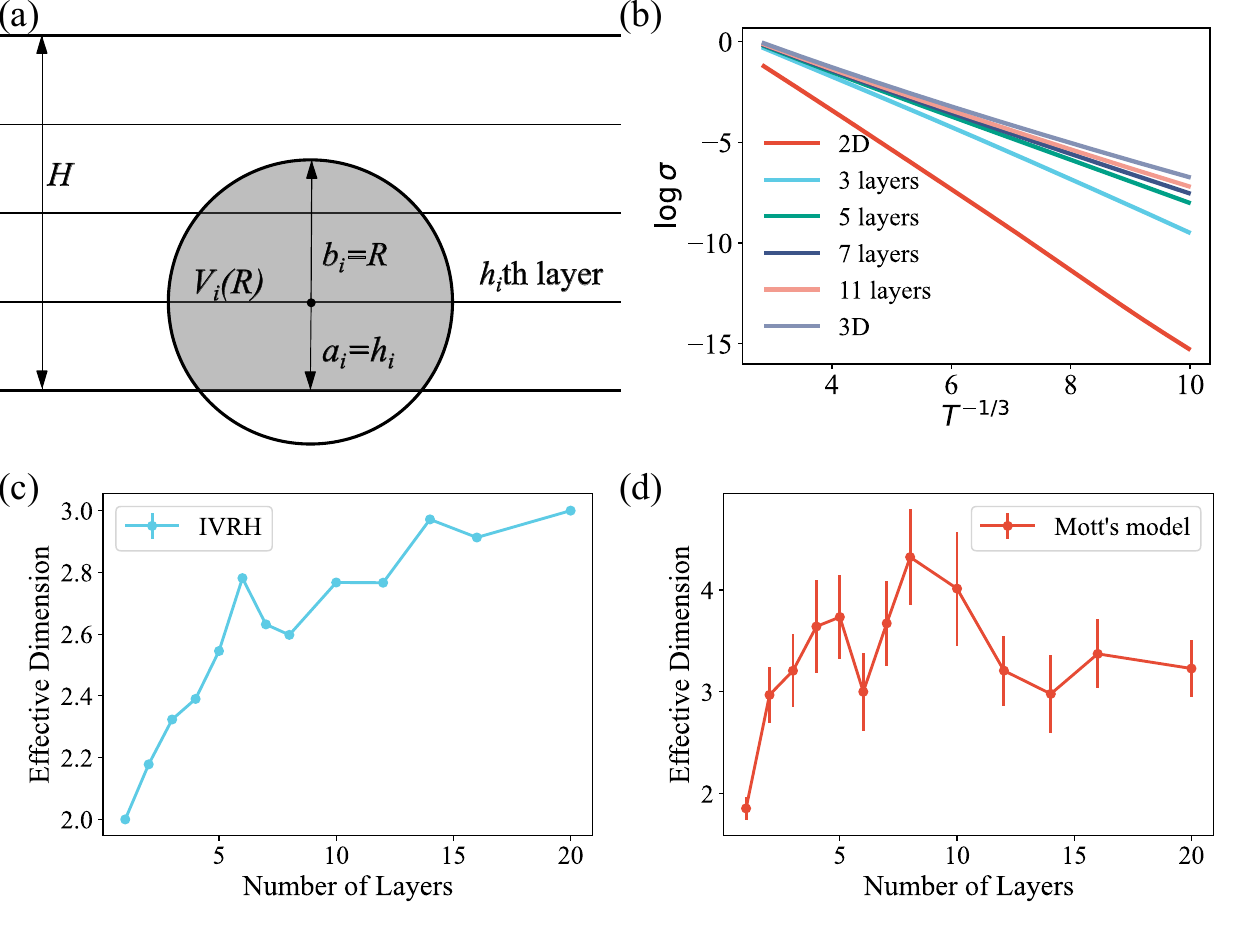}
    \caption{The application of IVRH in multi-layer systems and test with Monte Carlo results (a) Schematic of hopping geometry in multilayer systems. Here, the center of the sphere is located at the $h_i$th layer. The lower part of the sphere traverses the system boundary, thus $a_i=\min(h_i, R)=h_i$. The upper part is inside the system, so $b_i=\min(H-h_i, R)=R$. The shaded part corresponds to the effective hopping volume. (b) Temperature dependence of conductivity for various numbers of layers given by IVRH model. Parameters: $\alpha = \beta = D_0 = 1$.  (c)(d)Based on Monte Carlo simulation, the effective dimension of multi-layer systems predicted by the (c) IVRH model and (d) Mott's model, respectively.}
    \label{fig3}
\end{figure*}


Based on preliminary simulation results, we find that the linear regime is wider at higher temperatures. In order to fully take advantage of this property, we accept a scheme that adjusts the field range according to the temperature during the Monte Carlo simulation, as shown in Fig.\ref{fig2}(a). 
For lower temperatures, we choose a narrower field range to ensure the linearity. As the temperature grows, given the same field intensity, the error becomes larger, so we adopt a wider field range, which makes the linear fitting more stable and efficiently reduces the error bar. 

Fig.\ref{fig2}(b) presents the Monte Carlo simulation results alongside the corresponding fitting curves obtained using the IVRH model. For the two-dimensional case, we simulate a system with $L=20$ and $H=1$, applying periodic boundary conditions in the in-plane directions. For the three-dimensional case, we use a $L=20$, $H=20$ lattice with periodic boundary conditions in all three spatial directions. The fitted values of the model parameter $\beta$ are found to be 1.06 $\pm\ 0.02$ for the 2D system and 6.68 $\pm\ 0.21$ for the 3D system.
Assuming that the parameter $\beta$ is universal for systems of the same dimensionality, we proceed to analyze experimental conductivity data in disordered materials using these fitted $\beta$ values.

\begin{figure*}[]
    \centering
    \includegraphics[width=0.9\linewidth]{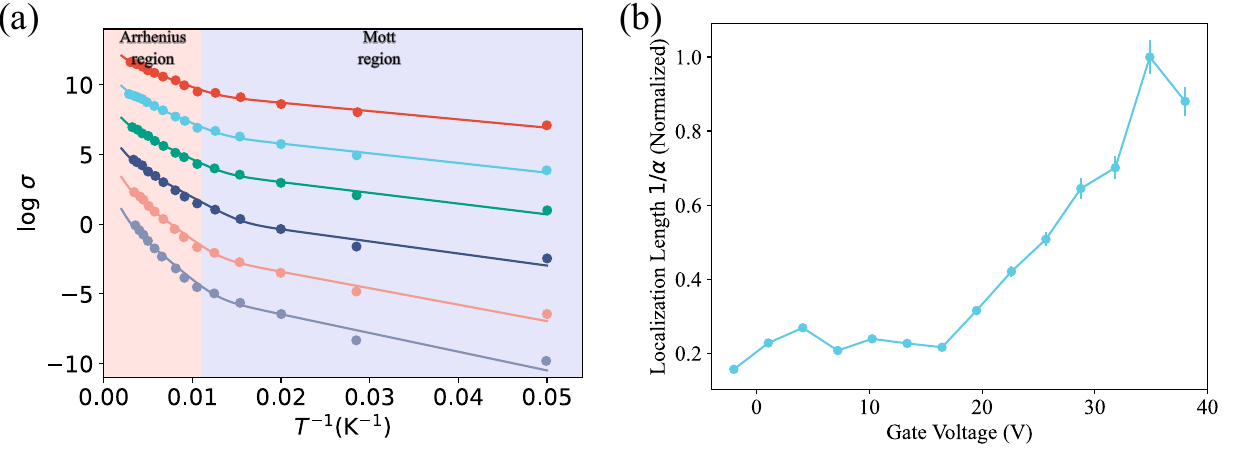}
    \caption{Application of IVRH to experimental measurements. \textbf{(a)} The temperature dependence of conductivity in a single-layer MoS$_2$ device. The data (dots) are fitted by the IVRH model (lines). 
    In the high-temperature (red) regime, the conductivity exhibits Arrhenius behavior. As the temperature is reduced, a crossover occurs into a low-temperature (blue) regime, where transport is governed by VRH and the conductivity follows Mott’s law.
    From top to bottom, the carrier density equals to 25, 20, 15, 10, 1$\times10^{11}$cm$^{-2}$, respectively. Experimental data are extracted from ref.\cite{qiu2013hopping}. \textbf{(b)} The value and fitted error bar of localization length under different gate voltages in a monolayer WS$_2$ device. Experimental data are extracted from ref.\cite{ovchinnikov2014electrical}.}
    \label{fig4}
\end{figure*}

\subsection{Extension to multi-layer system}

A merit of our IVRH model is that it can be readily extended to multi-layer materials, beyond pure two-dimensional or three-dimensional systems. 
The key step of this extension is to determine the appropriate form of the volume function $V(R)$ in Eq.\eqref{eq4}, which must account for confinement in the out-of-plane direction. \revise{Fig.\ref{fig3}(a) illustrates the effect of out-of-plane boundary on the hopping process. Specifically, when the hopping sphere traverses the boundary, its outer region (unshaded), which lacks accessible hopping sites, no longer contributes to the hopping behavior.} To quantify this, we define the effective volume of the hopping sphere centered at the $i$th layer as the volume of the hopping sphere inside the system:
\begin{equation}
    \begin{aligned}
        V_i(R)=\pi[a_i(R^2-\frac{a_i^2}{3})+b_i(R^2-\frac{b_i^2}{3})]
    \end{aligned}.
\end{equation}
where $a_i=\min(h_i,R),b_i=\min(H-h_i,R)$, and $h_i$ denotes the height of the $i$th layer in a system with $H$ total layers. This expression accounts for the truncation of the hopping sphere by the system boundaries.

Then, to describe the multi-layer system, we compute the layer-averaged hopping volume:
\begin{equation}
\bar V(R)=\frac{1}{H}\sum\limits_i^H V_i(R)    \label{c3eq3}
\end{equation}
Substituting $\bar{V}(R)$ into Eq.\eqref{e2} and integrating over $R$ yields the temperature-dependent conductivity $\sigma(T)$ for systems with various layer numbers $H$. The curves given by IVRH model for different $H$ are shown in Fig.\ref{fig3}(b). The $\log\sigma$–$T$ curves with different layer numbers interpolate the 2D and 3D situation. When $H > 10$, the system behavior closely converges to that of a 3D system with periodic boundary.

We further conduct Monte Carlo simulations for systems with various layer numbers and extract fitting parameters using both the IVRH model (Eq.\eqref{eq4}) and Mott model ($\log\sigma=BT^c+A$). In order to fairly compare these two models, we define the effective dimension of Mott's model and IVRH separately. For Mott's model, the power index $c$ equals $-1/(d+1)$ when $d$ is an integer. Therefore we define the effective dimension for Mott's model, $d^*_{M}$ as:
\begin{equation}
    d^*_{M}=-\frac{1}{c}+1
\end{equation}

As for the IVRH model, we define its effective dimension $d^*_{I}$ by normalizing the fitted parameter $\beta$ to $2\leq \beta \leq 3$:
\begin{equation}
    d^*_{I}=\frac{\beta-\beta_{2D}}{\beta_{3D}-\beta_{2D}}+2
\end{equation}
The results are shown in Fig.\ref{fig3}(c)(d). Surprisingly, when dealing with multi-layer systems, the IVRH model exhibits a much smoother behavior and has a negligible error bar, indicating the robustness and stability of the IVRH model. In contrast, the large uncertainties in $d^*_{M}$ limit the interpretability of Mott's model. This highlights a significant advantage of IVRH over traditional Mott-law fitting, particularly in layered systems and intermediate-dimensional systems.

\subsection{Analyses on experimental data}

After testing IVRH on the 2D, 3D and multi-layer systems with Monte Carlo simulation, we move on to investigate typical experimental observations. Fig.\ref{fig4}(a) shows the temperature dependence of the conductivity under different carrier densities measured with a single-layer MoS$_2$ device \cite{qiu2013hopping}. At high temperature, it features the Arrhenius behavior, while at low temperature it shows the VRH behavior. In the analysis of ref.\cite{qiu2013hopping}, these two regimes are manually separated and fitted individually. 
Here, we use IVRH to provide a unified description of the whole range, including the crossover region. 
As shown in Fig. \ref{fig4}(a), IVRH successfully recovers the experimental results. All the data points across the whole temperature range are utilized to determine the fitting curve, thus removing the ambiguity of manually identifying the crossover region and fitting the Arrhenius and Mott regions separately. The fitting results indicate that the temperature ranges associated with Arrhenius and Mott behaviors remain largely unchanged as the carrier density is varied, suggesting a relatively stable crossover regime.

To further highlight that our theoretical model naturally captures the physics behind experimental data, we can examine the transition upon tuning the gate voltage, during which a crossover between insulating and metallic states is observed in monolayer WS$_2$ \cite{ovchinnikov2014electrical}.
In Fig.\ref{fig4}(b), we plot the  value of localization length as a function of the gate voltage. At low gate voltage, the localization length is small; thus, the system is localized and exhibits insulating features. As the gate voltage increases, the localization length is larger, which means that the system is gradually delocalized. 

\section{Conclusion}
To conclude, in this work we have introduced the integral variable range hopping model, a physically grounded and computationally robust framework for analyzing hopping transport in disordered systems across universal dimensions. 
By integrating over hopping distances and parametrization of the empirical temperature exponent $\beta$, IVRH offers improved fitting stability and clearer physical interpretation compared to the vanilla Mott's variable range hopping model. IVRH also provides a natural description for the transition between nearest hopping to variable range hopping, and exhibits a unique advantage in interpreting experimental observations. Moreover, our model is readily extendable to multi-layer systems through the geometric modification of $V(R)$, enabling systematic studies across dimensional crossover regimes. This approach opens a new avenue for interpreting experimental data on amorphous semiconductors and disordered thin films with improved precision and reduced ambiguity. It still requires further research to consider the Coulomb interaction in IVRH model, which in principle can be done by modifying the density of states near the Fermi surface.

\begin{acknowledgments}
This work was supported by the National Science Foundation of China under Grant No. 52541026.
We are grateful for computational resources provided by the High Performance Computing Platform of Peking University.
\end{acknowledgments}

\bibliography{ref}
\bibliographystyle{unsrt}
\end{document}